\documentclass[]{spie}  

\usepackage[]{graphicx}

\def\deg{\hbox{$^\circ$}}

\def\lae{\mathrel{\raise .4ex\hbox{\rlap{$<$}\lower 1.2ex\hbox{$\sim$}}}}
\def\gae{\mathrel{\raise .4ex\hbox{\rlap{$>$}\lower 1.2ex\hbox{$\sim$}}}}

\title{Progress toward a Soft X-ray Polarimeter} 

\author{Herman L.\ Marshall\supit{a},
Norbert S.\ Schulz\supit{a},
Brian Remlinger\supit{a}, 
Eric S.\ Gentry\supit{a},
David L. Windt\supit{b},
Eric M.\ Gullikson\supit{c}
\skiplinehalf
\supit{a}MIT Kavli Institute, Cambridge, MA, USA 02139\\
\supit{b}Reflective X-ray Optics, 1361 Amsterdam Ave, Suite 3B, New York, NY, USA 10027\\
\supit{c}Lawrence Berkeley National Lab, 1 Cyclotron Rd., Bldg. 2R0400, Berkeley, CA, USA 94720
}

\authorinfo{Further author information: (Send correspondence to H.L.M.)\\H.L.M.: hermanm@space.mit.edu,
Telephone: 1 617 253 8573} 

\begin{document} 
  \maketitle 

\begin{abstract}
We are developing instrumentation for a telescope design capable of measuring linear X-ray polarization
over a broad-band using conventional spectroscopic optics.  Multilayer-coated mirrors are key to this
approach, being used as Bragg reflectors at the Brewster angle.
By laterally grading the multilayer mirrors and matching to the dispersion of a spectrometer,
one may take advantage of high multilayer reflectivities and achieve modulation factors
over 50\% over the entire 0.2-0.8 keV band.
We present progress on laboratory work to demonstrate the capabilities of an existing
laterally graded multilayer coated mirror pair. 
We also present plans for a suborbital rocket experiment designed to detect a polarization
level of 12-17\% for an active galactic nucleus in the 0.1-1.0 keV band.

\end{abstract}


\keywords{X-ray, polarimeter, astronomy, multilayer, mirror, grating}

\section{INTRODUCTION}
\label{sec:intro}  

We continue our investigation and laboratory work to develop a soft
X-ray polarimeter based on Bragg reflection from multilayer-coated optics.
Marshall (2007\cite{2007SPIE.6688E..31M})
described a method using transmission gratings
to disperse the incoming X-rays so that the dispersion
is matched to laterally graded multilayer (ML) coated reflectors.  
An extension of this approach was suggested by
Marshall (2008\cite{2008SPIE.7011E..63M})
that can be used with larger missions
such as the AXSIO or AEGIS.
Some potential scientific investigations that
would be possible with a soft X-ray polarimeter
were described earlier and include testing the
synchrotron nature of quasar jet emission and
models of neutron star atmospheres\cite{plexas,2010SPIE.7732E..12M}.

The laboratory work was initiated in order to test
prototype optics for eventual use in a flight design.
The work started with simple measurements after
recommissioning the 17-m X-ray beamline at MIT's
building NE80.  Murphy et al. (2010\cite{2010SPIE.7732E..108M})
showed early results and described most of the system
in detail.  Here, we show results after improving alignment,
resulting in a uniform, monochromatic beam, and
the first successful polarization test.
We also
describe work to develop new ML coatings and
laterally graded ML coated mirrors (LGMLs) and the next steps
in our lab development.

A description of a design for a suborbital rocket
flight is given in \S\ref{sec:instr}.
The experiment's  minimum detectable polarization (MDP) is
expected to be 6.5\% when observing a bright blazar such
as Mk 421.

\section{The MIT Polarimetry Beamline}

We have recently recommissioned the
X-ray grating evaluation facility (X-GEF), a 17 m beamline that was developed
for testing transmission gratings fabricated at MIT for
the {\it Chandra} project\cite{1994SPIE.2280..257D}.
The project development is proceeding in four distinct phases, of which
two have been completed.  In Phase I, we set up the polarized X-ray
source at one energy (0.525 keV) and aligned it so that the beam was
uniform at the detector and its intensity did not vary
significantly with rotation angle.  In Phase II, we added a ML coated mirror
to the detector end of the system and reoriented the
detector to face 90$\deg$ to the beamline and demonstrated that
the source produced nearly 100\% polarized X-rays.

We are now in Phase III, where we are replacing the source ML
coated mirror with a LGML in order to polarize a range of input emission
lines from 0.2 to 0.8 keV.  In Phase IV, we will insert a diffraction grating
in order to disperse the polarized input onto an LGML in the detector
chamber.  At the end of Phase IV, we will be ready to test LGMLs with
improved reflectivities and with larger ML coating periods in order to
demonstrate that they can be used in a flight system, such as the
one described in \S\ref{sec:instr}.

\subsection{Polarimetry Beamline Phase I}

\begin{figure}
   \centering
   \includegraphics[width=9cm]{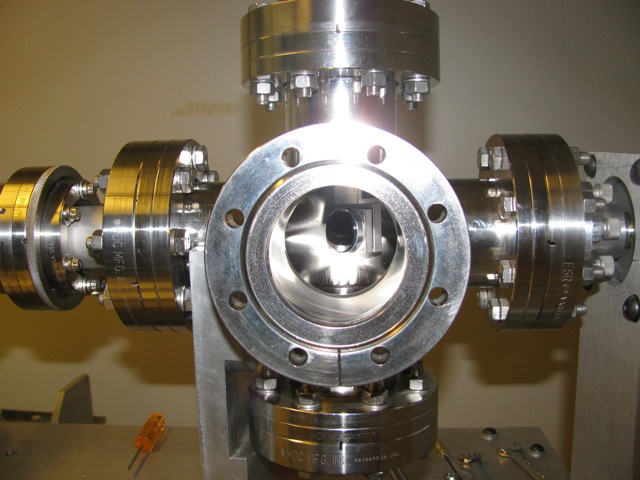}
   \includegraphics[width=7cm]{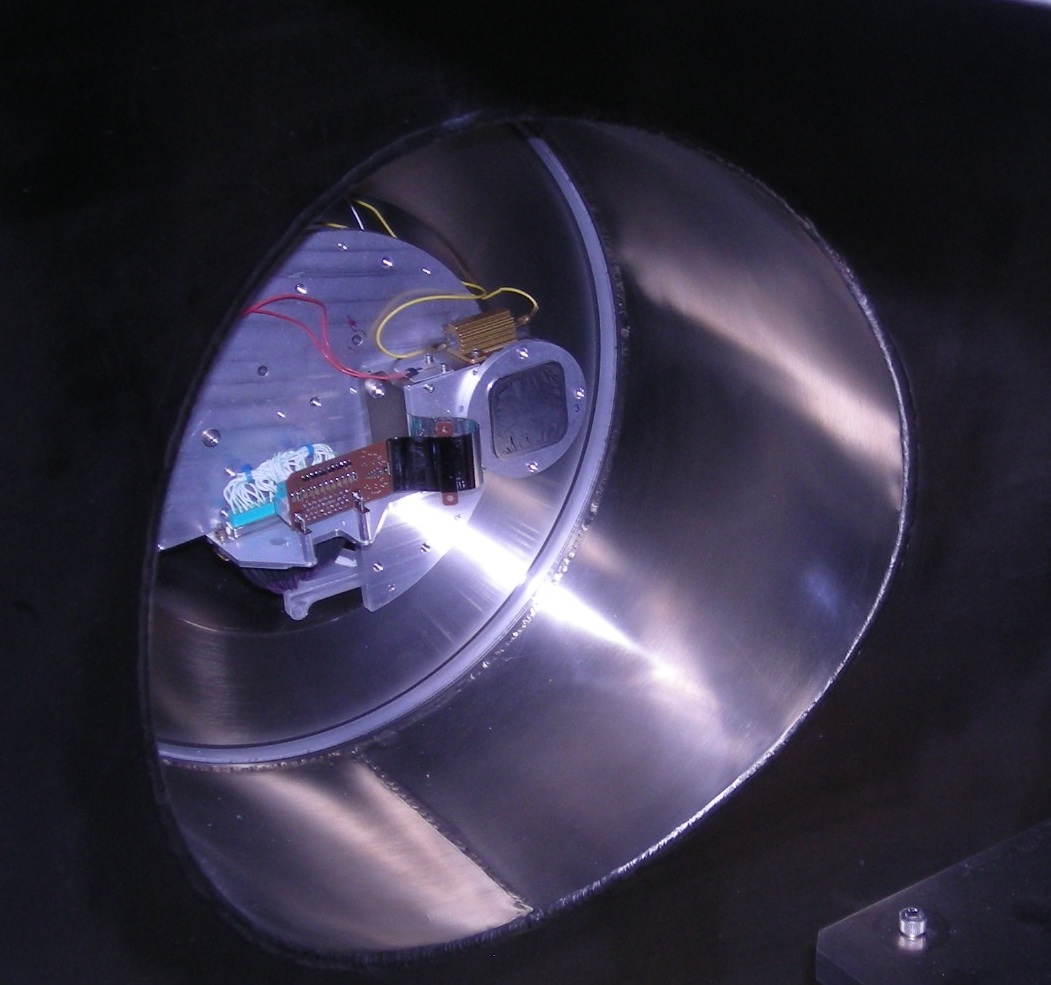}
 \caption{
 Hardware in the polarimetry beamline for Phases I and II.
 {\it Left:} A single-period ML coated mirror is mounted at 45$\deg$ to the beamline
  (at left), as viewed from the flange to which the X-ray source is mounted.  The
  mirror is about 25 mm in diameter and is attached to an Al mount with adjustment
  screws to rotate it about two axes: vertical and horizontal.
  The rotatable vacuum flange is shown at left and the motor that rotates the
  source and mirror chambers is attached at right.  The source rotation phase
  angle is defined to be zero at the orientation shown in
  the picture and 90$\deg$ when the X-ray
  source has been rotated to be above the beamline.
 {\it Right:} The CCD detector, mounted on a 200 mm flange, as viewed
 from the detector chamber, about 17 m from the X-ray source.
}
\label{fig:phase1hw}
\end{figure}

\begin{figure}
  \centering
   \includegraphics[width=\columnwidth]{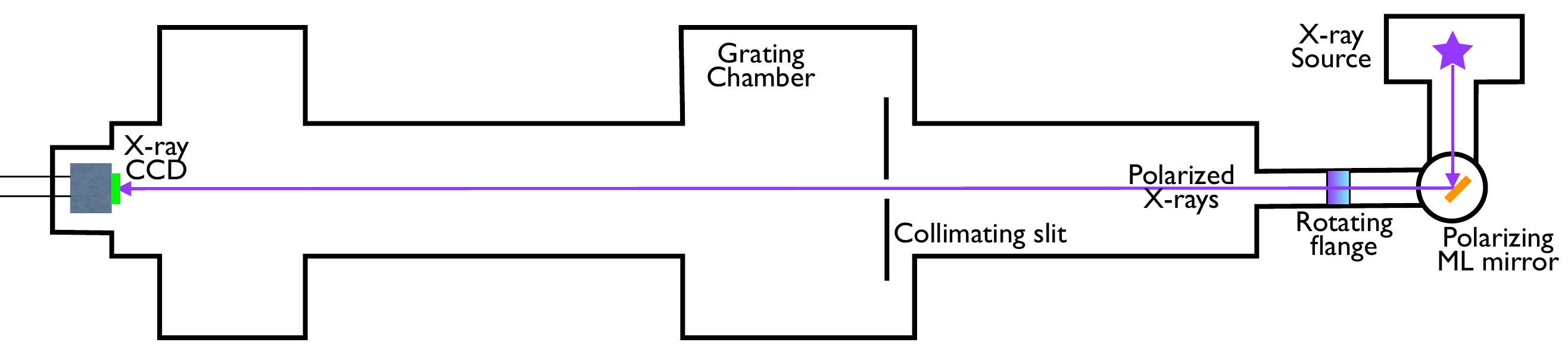}
 \caption{
 Schematic of the polarimetry beamline in Phase I.  The polarization E-vector is oriented
 perpendicular to the plane of the drawing, for this orientation of the rotating flange.
 The equipment to the right of the flange is rotated about the beamline axis under
 computer control by a motor that can also hold the source
 in place at a range of angles from $-35\deg$ to $+130\deg$, where angles are
 defined using the right-hand rule pointing to the detector.  The diagram shows
 the configuration at an angle of $+90\deg$.  The collimating slit is a 25 mm
 square aperture in this phase.
}
\label{fig:schematic1}
\end{figure}

With MKI technology development funding, we 
adapted the source to produce polarized X-rays at the O-K$\alpha$
line (0.525 keV)\cite{2010SPIE.7732E..108M}.
A five-way chamber was added to house the Polarized Source MultiLayer (PSML) mirror
(see Fig.~\ref{fig:phase1hw}).
The mirror and a twin were provided by Reflective X-ray Optics (RXO), with a coating consisting
of 200 layers of 5.04\AA\ of W alternating with 11.76 \AA\ of B$_4$C.
The wavelength, $\lambda$, of the Bragg peak for a periodic ML coating is given by
$\lambda = 2 D \sin \theta$, where
$D = 16.80$\AA\ is the average layer thickness, and $\theta = 45\deg$ is the
graze angle (measured from the surface).
The source is mounted to this chamber
at $90\deg$ to the existing beamline, illuminating the ML mirror
at a 45$\deg$ angle to the X-ray source and the output port.
A rotatable flange connects the
PSML chamber's output port to the vacuum pipe so that the polarization
vector can be rotated through an angle of about 160$\deg$.
For this phase, the collimator plate is set to an 25 mm square aperture, which
provides a 50 mm wide illumination pattern at the detector.
A front-side illuminated CCD spare from the {\em Suzaku} project \cite{2004SPIE.5501..385L}
was installed on the output port of the detector chamber (see Fig.~\ref{fig:phase1hw}).
The CCD has 1024$^2$ 24 $\mu$ square pixels and is read out by prototype electronics
connected to a Sun workstation.
See Murphy et al. (2010\cite{2010SPIE.7732E..108M})
for early details and pictures of the soft X-ray polarization laboratory.
The schematic is shown in Fig.~\ref{fig:schematic1}.

The bulk of the work in Phase I was in stabilizing the PSML chamber
and aligning the beam so that there would be minimal variation of
the count rate in the detector.  Fig.~\ref{fig:phase1hw} shows the mirror
chamber before a hold-down bracket was installed on the left support
bracket in order to prevent lifting of the chamber as the source was
rotated about the beamline.  The mounts on both sides are secured
to the table, which is epoxied to the floor.  The ML mirror mount has
adjustment screws for two axes of rotation and is attached to the
shaft of a rotatable manipulator via a small bracket.  Upon stabilizing
the system, measurements were consistently repeatable as the
source was rotated.

Alignment was achieved by setting up a laser in the detector chamber
about 16 m from the X-ray source and pointed through the square
collimating aperture to the polarizing mirror, thus defining the beamline
optical axis.  The mirror readily reflects
laser light, so it was rotated to be perpendicular to the laser beam,
in order to reflect the laser beam back to the collimating aperture.
The mirror chamber mounts (see Fig.~\ref{fig:phase1hw}, left) were
adjusted vertically and horizontally and the mirror mount was adjusted
to ensure that the reflected laser beam was centered on the
collimating aperture, even as the mirror chamber was rotated.  In
this manner, the mirror's axis was aligned to the beamline axis
to within 1 cm along an 8.5 m length, for an accuracy of
about $\pm 0.001$ radians.

The rotation angle of the mirror
about the manipulator's shaft was set using an alignment laser
installed near the source, downstream of the rotating flange.
This laser is oriented perpendicular to the beamline, illuminating
an insertable prism that provides oppositely directed output beams.
The direction of the beam along the beamline is controlled with a
micrometer to pass through the collimating aperture in the grating
chamber and illuminate the CCD in the detector chamber.  The
opposite beam reflects off of the source ML mirror to place a small
spot on the X-ray source anode, viewable through a viewport
at the source.  The source anode used
for these tests consisted of a Cu anode with a small sapphire
(Al$_2$O$_3$) disk glued into the center of the target face.  The
anode is oriented at 45$\deg$ to both the output and view ports, so
the laser light reflects off of the sapphire disk and can be viewed
on a translucent screen attached to the viewport.  Marking the screen
ensures that the mirror orientation can be recovered upon
disassembly and subsequent reassembly.

 \begin{figure}
  \centering
   \includegraphics[width=6cm]{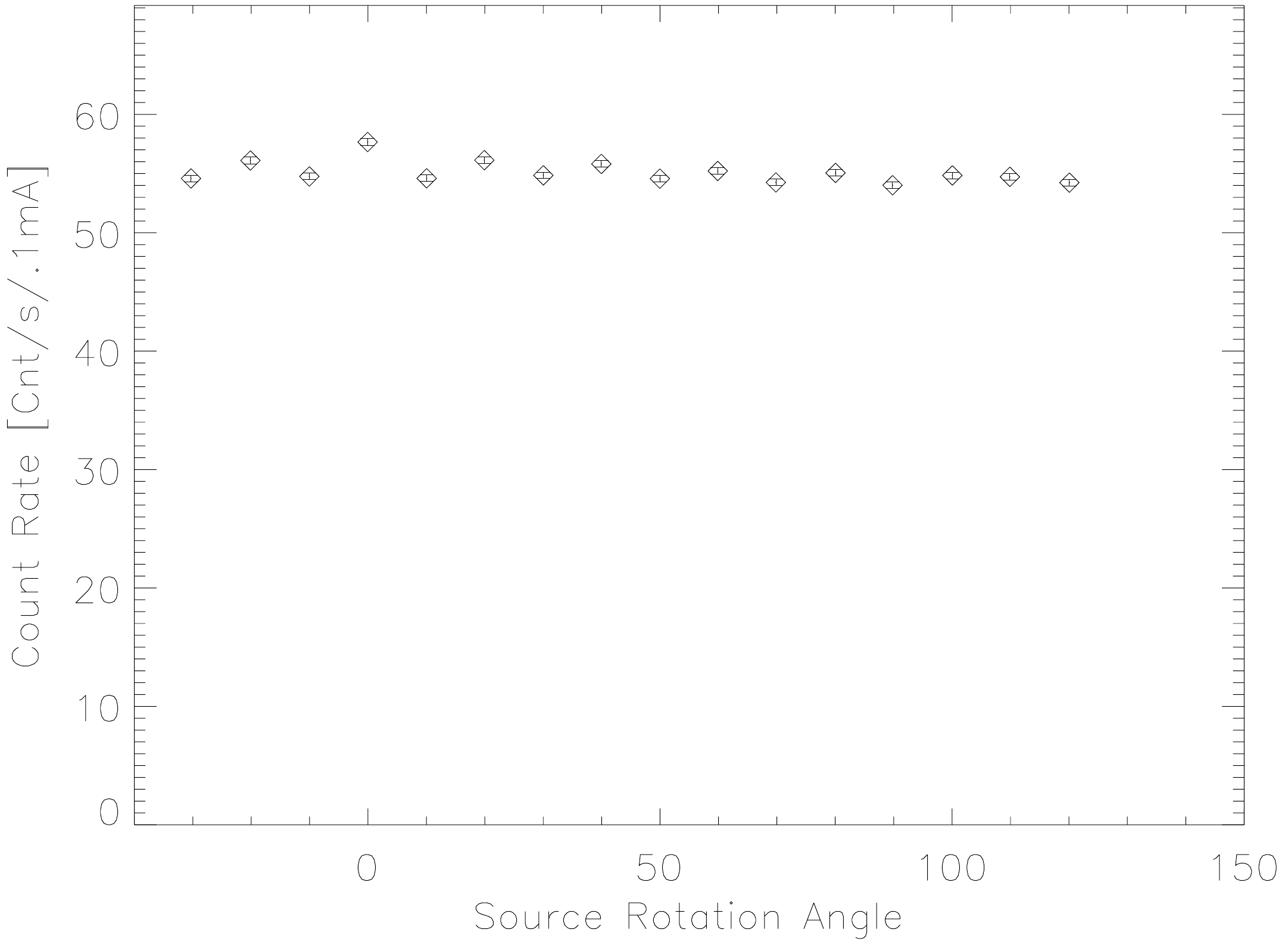}
   \includegraphics[width=4.5cm]{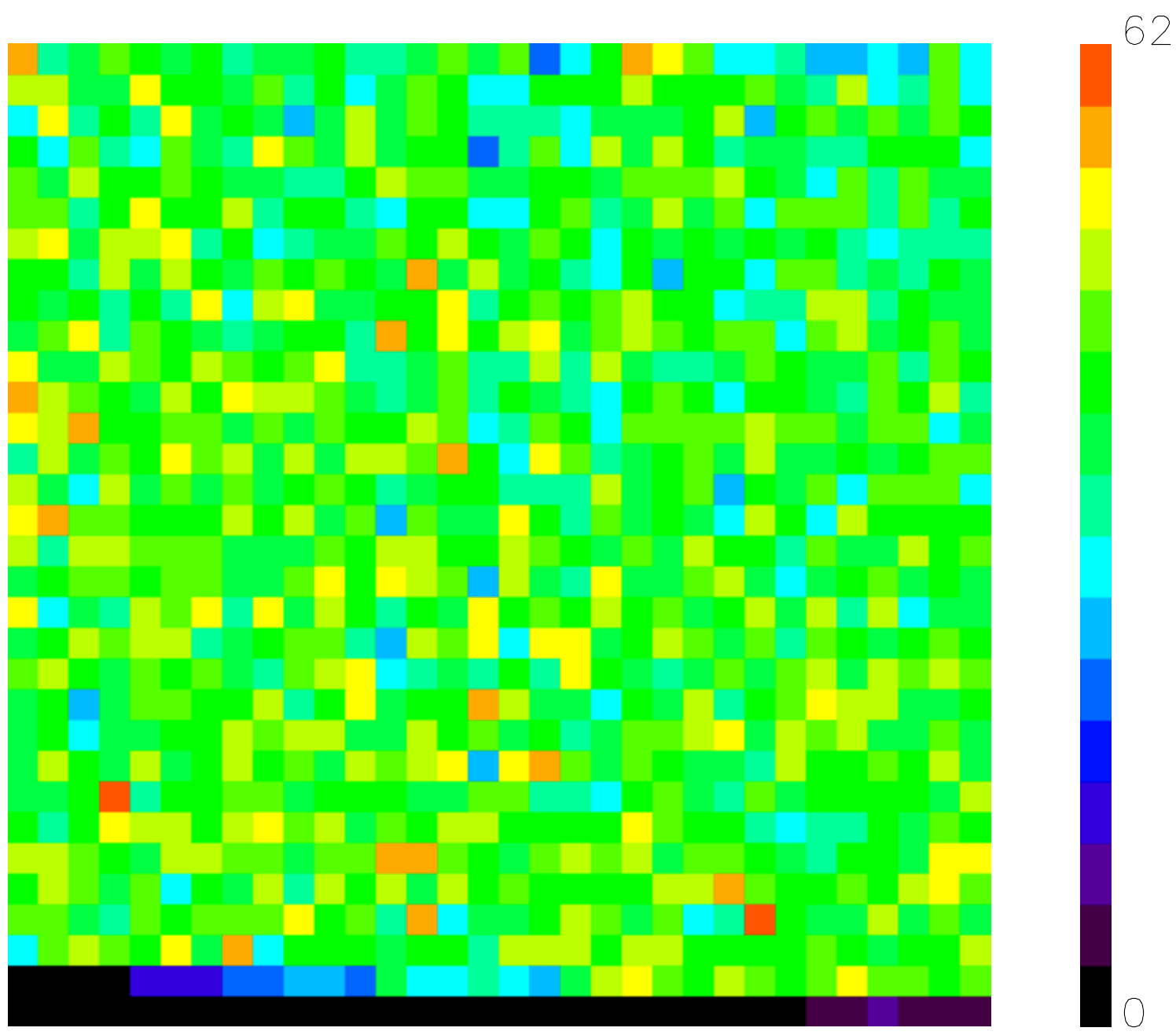}
   \includegraphics[width=6cm]{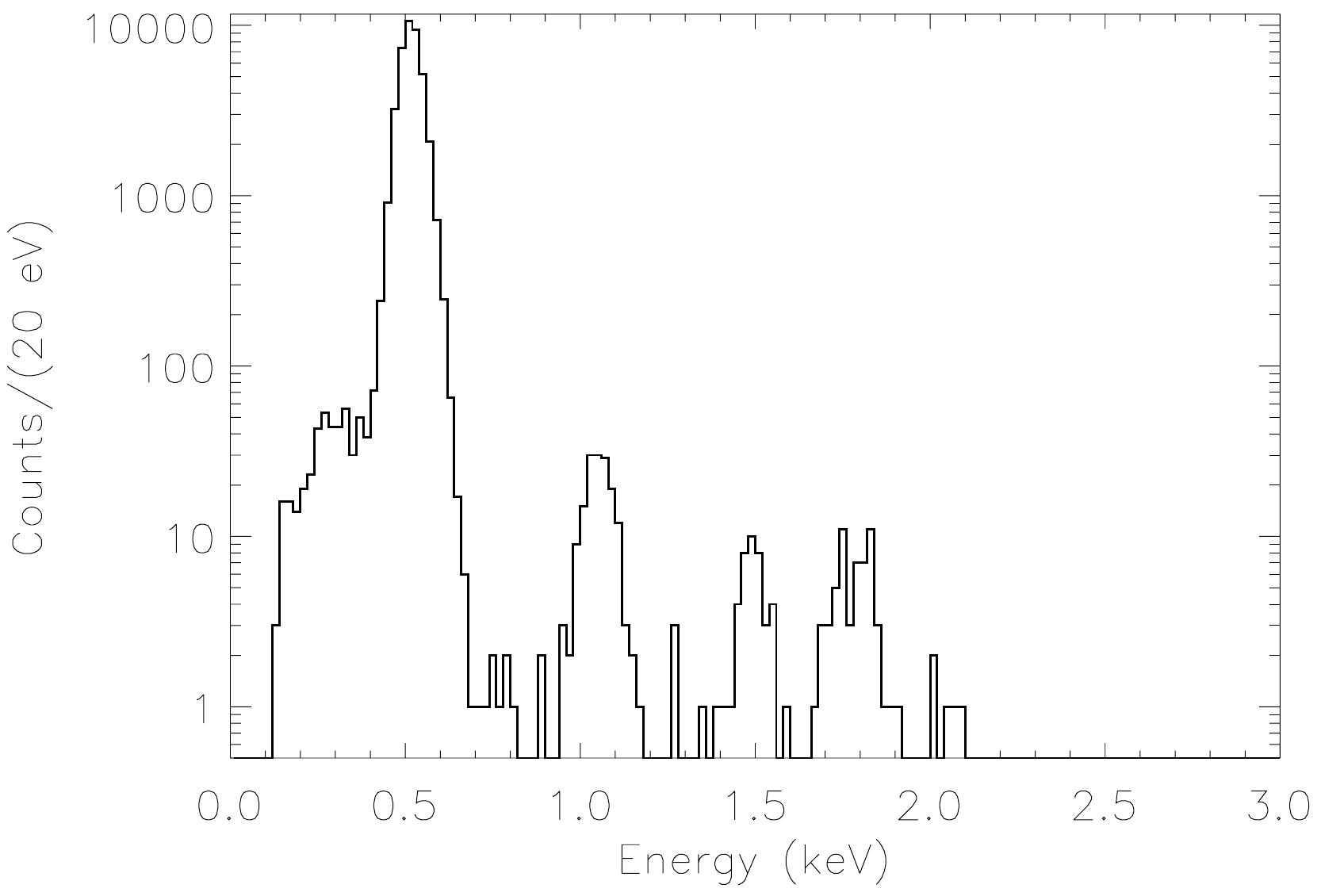}
 \caption{
 Results from Phase I operations of the polarimetry beamline, where the CCD
  receives the polarized beam directly (see Fig.~\ref{fig:schematic1}).
 {\it Left:} The CCD count rate per 0.1 mA as a function of polarization rotation
 angle.
 The source was operated at a voltage of 5 kV and a beam current of 0.3 mA, for
   240 s exposures on 16 July 2013.
 The beam is uniform with system rotation to better than 5\%.
 {\it Center:} CCD image of one of the observations, taken at a rotation angle
 of 50$\deg$.  The variation of the counts in 32 $\times$ 32 pixel regions is
 only 5\% greater than expected from Poisson counting statistics, indicating
 that the beam is uniform over scales from 0.8 mm to 2.5 cm.
 {\it Right:} Pulse distribution from the observation at 50$\deg$.  Over 95\% of all
 events are in the O-K peak, as expected.  The other features are a tail due to hot
 pixels below 0.4 keV and fluorescence from elements in the multilayer,
 its substrate, and the mirror holder.
}
\label{fig:uniform}
\end{figure}

Figure \ref{fig:uniform} shows results from tests after alignment.
The source was operated at a voltage of 5 kV and a beam current of 0.3 mA, for
240 s exposures on 16 July 2013.
Measurements were obtained at 20$\deg$ steps rotating the system
in one direction, then shifting 10$\deg$ and stepping 20$\deg$
in the reverse direction.
The count rate is consistent to within 5\% with rotation of the
source and mirror chamber, as shown in the left panel.
The middle panel shows that the CCD image is uniform to within 5\%
on small scales.
The right panel shows an example of the pulse height distribution
measured by the CCD for one exposure.  Over 95\% of all
events are in the O-K peak, as expected.  The other features are a tail due to hot
pixels below 0.4 keV combined with B-K and C-K fluorescence photons,
pulse pileup from two events in a single frame at
1.05 keV, Al-K fluorescence near 1.5 keV, and W-M and Si-K
fluorescence near 1.8 keV.  Fluorescence can result from bremsstrahlung photons
from the source penetrating the ML coating (comprised of C, B, and W) to
the substrate (containing Si) and illuminating the Al mirror holder.

\subsection{Polarimetry Beamline Phase II}

\begin{figure}
  \centering
   \includegraphics[width=\columnwidth]{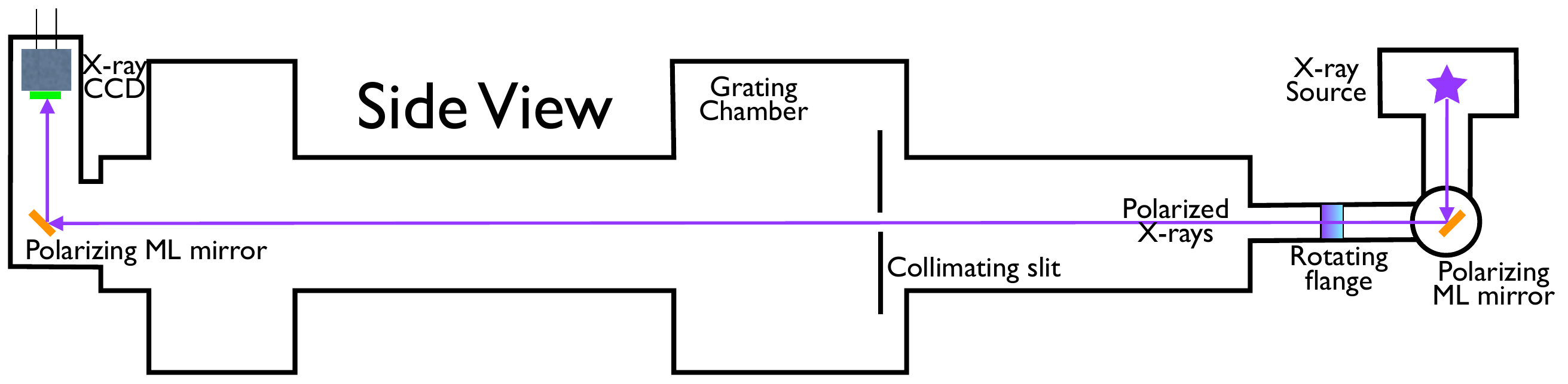}
 \caption{
 Schematic of the polarimetry beamline in Phase II.  The configuration is the same
 as in Phase I (Fig.~\ref{fig:schematic1}) except that a polarizing ML mirror has been added to detector
 end of the beamline and the CCD is reoriented to face it.
}
\label{fig:schematic2}
\end{figure}

\begin{figure}
  \centering
  \includegraphics[width=3.5cm, angle=0]{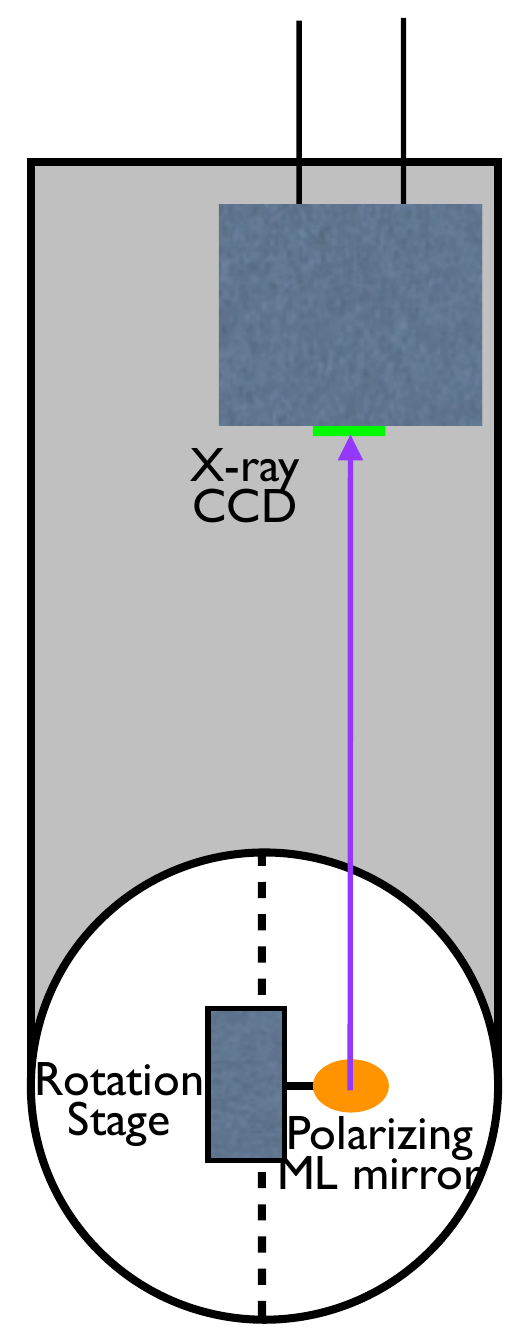}
   \includegraphics[width=6cm, angle=0]{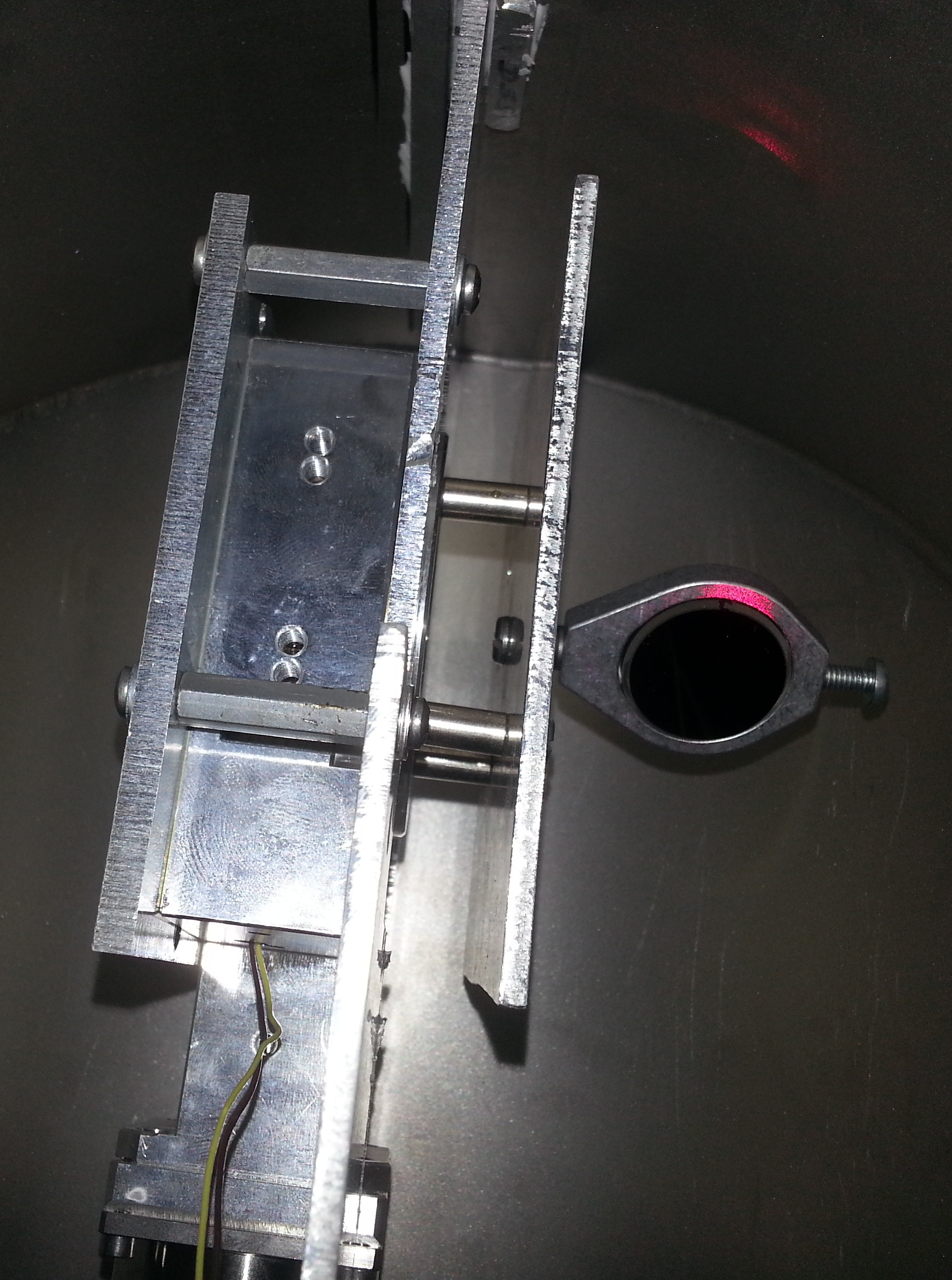}
   \includegraphics[width=6cm, angle=0]{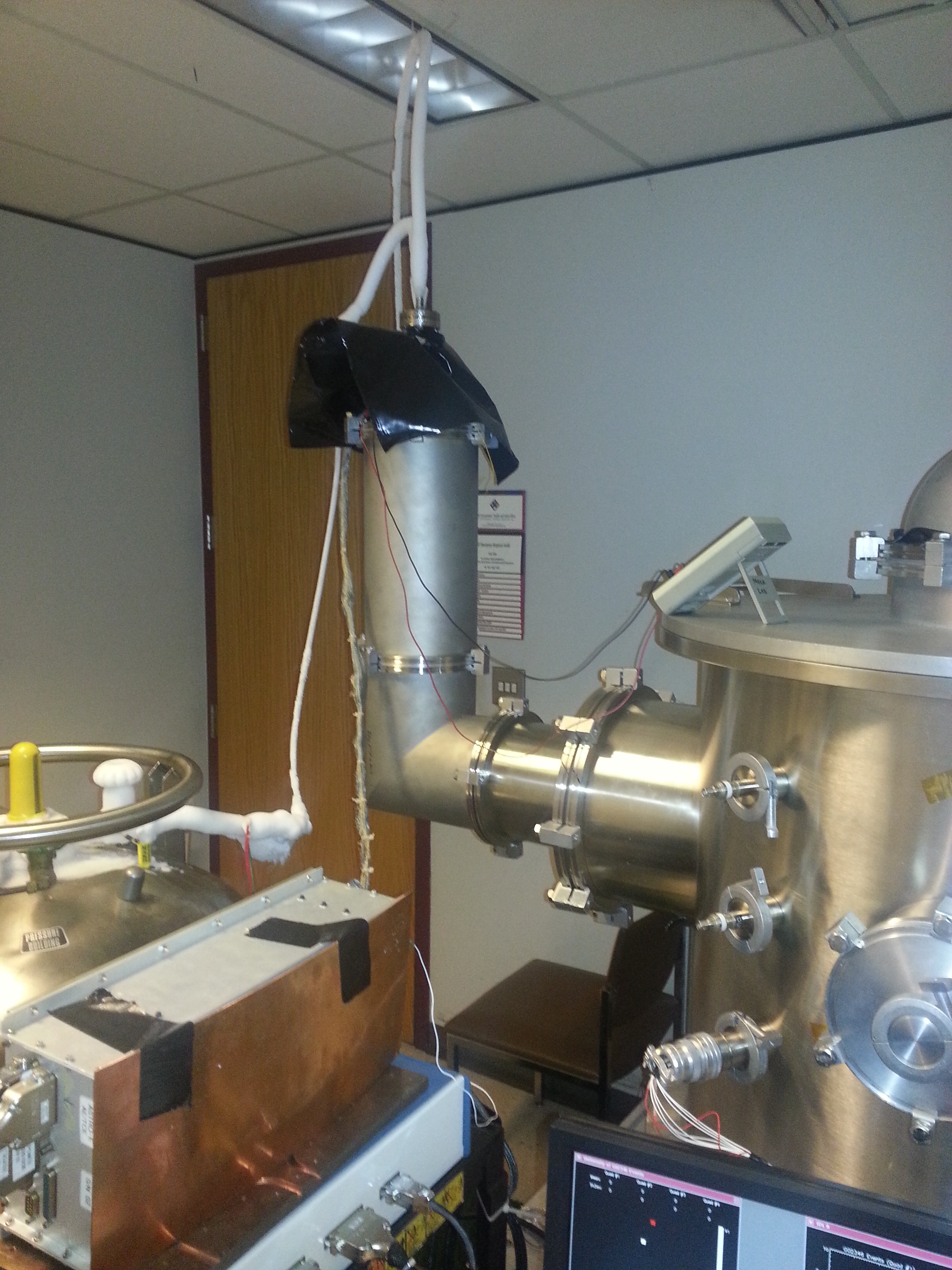}
 \caption{
 Hardware added to the polarimetry beamline for Phase II.
 {\it Left:} Schematic of the added equipment, as seen from the source along
 the beamline.
 {\it Middle:} A single-period ML coated mirror is mounted at 45$\deg$ to the beamline
  (from below), as viewed from the flange to which the detector is mounted.  The
  mirror is a twin of the source mirror (see Fig.~\ref{fig:phase1hw}, left) and
  is mounted to a rotational stage to allow rotation about the mount axis.
  The source alignment laser, shown illuminating part of the mirror mount,
  is used to position and align the mirror.
 {\it Right:} Exterior view of the revised detector end of the system.  The
 ML mirror is mounted in the elbow pipe and reflects X-rays to the detector
 mounted vertically at the top of the 200 mm diameter extension pipe.
 Also visible and used in Phase I
 are the detector chamber, liquid N$_2$ tank and frosted supply lines, the
 wrapped data and command
 wire bundle, a multimeter connected to a resistor on the CCD
 mount plate to measure its temperature, the CCD electronics boxes,
 and the screen of the computer that receives data from and sends commands to
 the CCD.
}
\label{fig:phase2hw}
\end{figure}

The objective of phase II was to demonstrate that source-mirror combination
produces 100\% polarized X-rays.  The schematic is shown in Fig.~\ref{fig:schematic2}
and is the same as for Phase I (Fig.~\ref{fig:schematic1}) except for the addition
of an elbow pipe in which a ML coated mirror is installed and oriented at 45$\deg$ to
the beam.  The mirror is a twin of the source mirror, so it is designed to reflect
O-K photons with s-polarization.  In the schematic, only those photons with polarizations
perpendicular to the plane of the drawing will be reflected.
Fig.~\ref{fig:phase2hw} shows pictures of the added hardware.

As in Phase II, substantial care was taken in mounting and aligning the ML mirror.
Alignment was achieved using the system alignment laser mounted in the source
end of the system.  When reflected off of the mirror, the beam was centered in the
vertical pipe along the plane perpendicular to the beam.  We also verified that the
reflected laser spot illuminated the CCD after it was attached to the vertical pipe
by using a camera held in the elbow pipe.

\begin{figure}
   \centering
   \includegraphics[width=8cm]{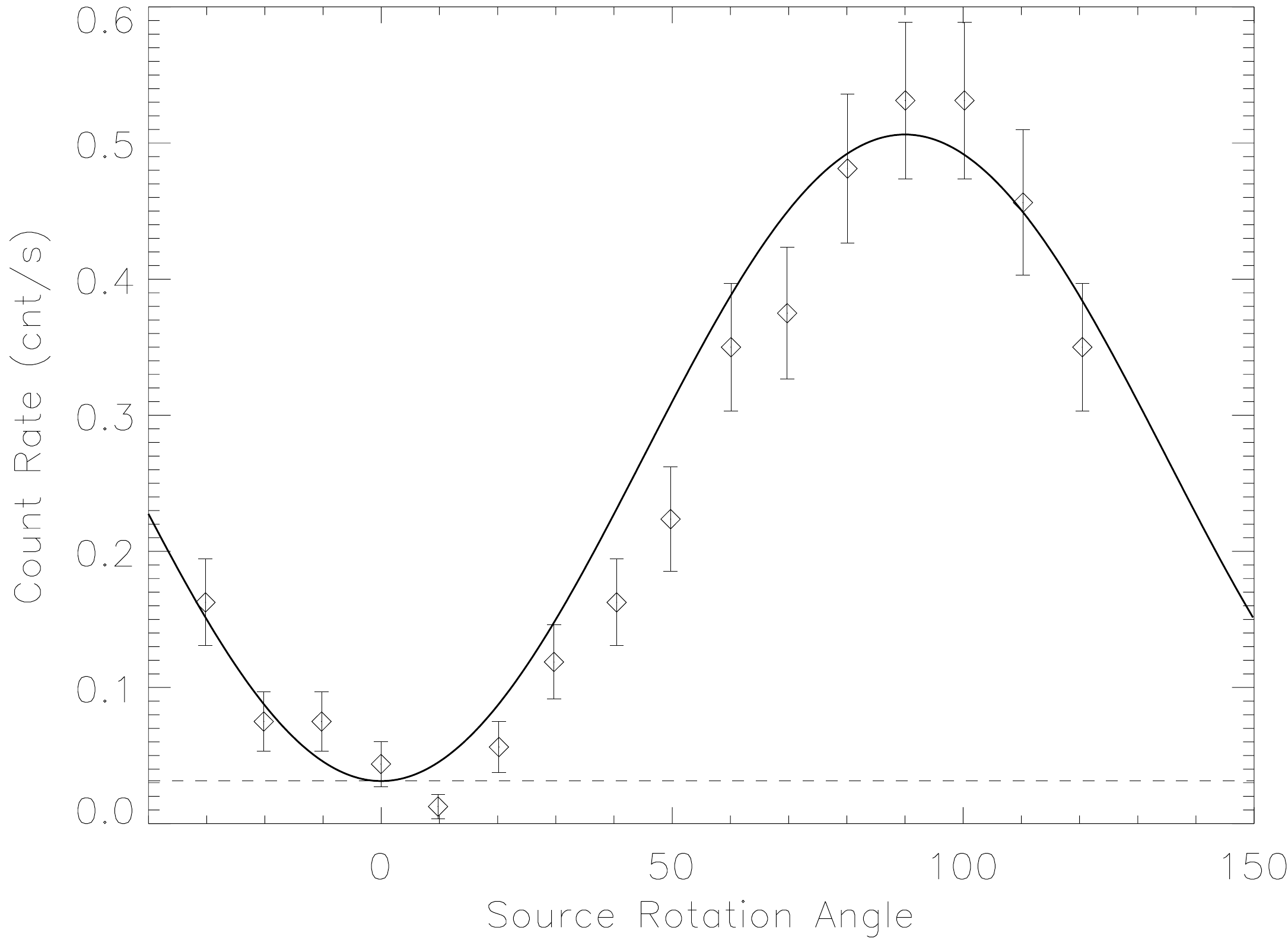}
   \includegraphics[width=8cm]{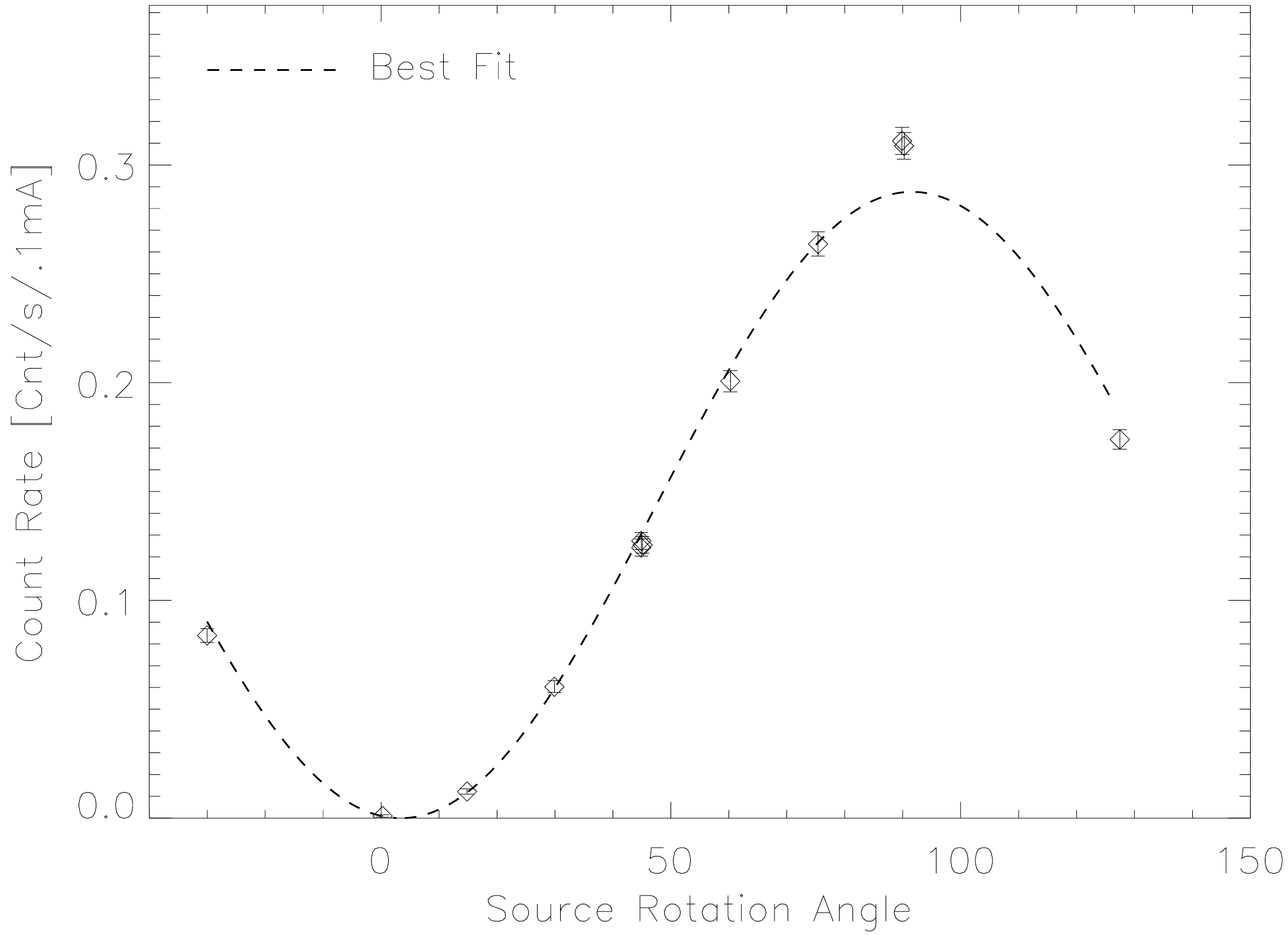}
 \caption{
 Results from Phase II operations of the polarimetry beamline, where
 the detector polarizing mirror is installed
 and the detector is 90$\deg$ to the beamline.  (see Fig.~\ref{fig:schematic2}).
 {\it Left:} The source was operated at a voltage of 5 kV and a beam current of 0.3 mA for
   160 s exposures on 3 August 2013.  The dashed line indicates the maximum possible
   dark contribution.
 {\it Right:} The source was operated at a voltage of 8 kV and a beam current of 0.5 mA for
   1600 s exposures on 5 August 2013.  The dark current is negligible, so the polarization model,
   a simple sine wave, is an excellent fit to the data.  The phase angle of the minimum is
   consistent with 0$\deg$, as expected.
}
\label{fig:result}
\end{figure}

Figure~\ref{fig:result} shows results from two different runs in this configuration.
In the first one (left panel), only 160 s exposures were obtained at each angle.  The
X-ray source was operated at a voltage of 5 kV and 0.3 mA, as was used
to obtain the uniformity results shown in Fig.~\ref{fig:uniform}.  The background
was not well determined for the first observation.
For the second run, the source voltage was raised to 8 kV and the current
raised to 0.5 mA in order to achieve near maximal source emission.
The background was very small,
so the modulation factor is consistent with the expectation that the beam is 100\%
linearly polarized.  As expected, the count rate null is found near 0$\deg$ and the
maximum is at 90$\deg$.  The rates are repeatable to within the counting statistics
of the measurements, even when several thousand counts were obtained.

\begin{figure}
   \centering
   \includegraphics[width=5.25cm]{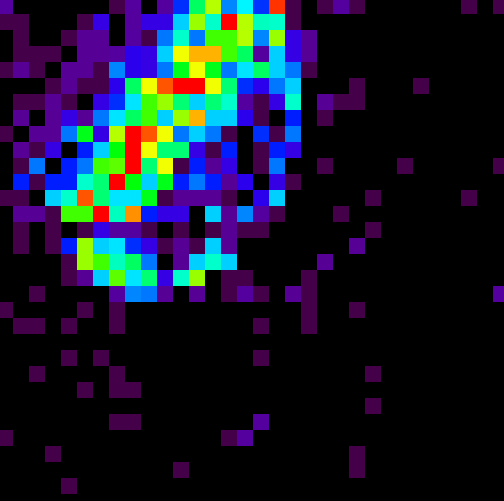}
   \includegraphics[width=5cm]{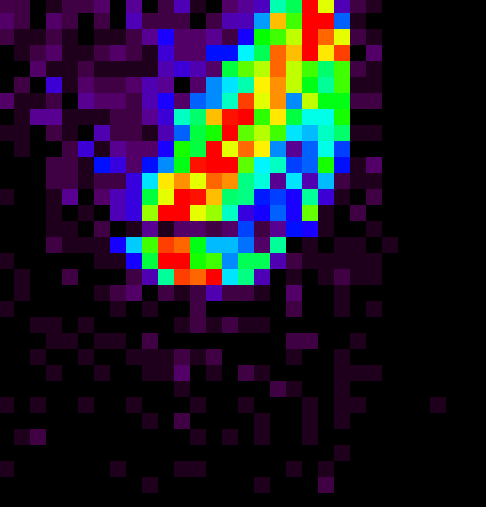}
   \includegraphics[width=5.2cm]{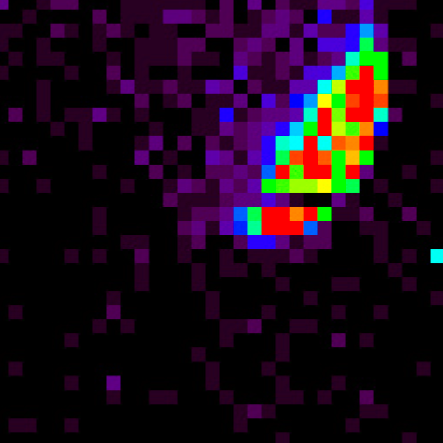}
 \caption{
 CCD images from Phase II operations of the polarimetry beamline.
 The $x$ coordinate increases in the direction of the X-ray source.
 The mirror and CCD are both 25 cm across but the mirror
 is tilted at 45$\deg$ to the normal of the CCD, so the
 image from a fully illuminated mirror would be 70\% of the width of the
 CCD.
 The detector polarizer mirror was rotated through a range of $0.6\deg$
 across the set of observations, both moving the image slightly and changing the
 image's beam pattern.  We deduce that the Bragg peak varies
 across the surface of the mirror by 1.0\% from center to edge.
}
\label{fig:images}
\end{figure}

Images from these tests showed that the mirror did not reflect all incident photons
to the detector.  Figure~\ref{fig:images} shows that the pattern of illumination on the CCD varied
with mirror rotation angle.  The pattern is caused by a combination of a nearly
monochromatic input beam, a narrow reflectivity curve, and a slight variation of the
ML period across the mirror.  The mirror, of diameter $d = 25$ mm at distance $L$ from
the source ML coated mirror,
subtends an angle of $\Delta \theta = d/L = 0.0015$ rad as seen
by the source mirror.  Differentiating the Bragg condition gives the spread of
the peak wavelength,
$\Delta \lambda = 2 D \cos \theta ~ \Delta \theta$, so the beam's
spectral resolution is $\Delta \lambda / \lambda =  \tan \theta ~ \Delta \theta = \Delta \theta$
for $\theta \approx 45\deg$.
Thus, the spectral width of the incident beam in our setup is
about $\Delta E = 0.0015 E_{\rm O-K} = 0.8$ eV.
The ML reflectivity curve is about 3 eV FWHM, so the input beam samples a small range
of the reflectivity curve.  When $D$ varies by 1\%, the Bragg peak changes by 5 eV,
where the reflectivity is less than 10\% of the peak.  In order to test this possible cause
of the nonuniformity, the ML mirror was rotated through
an angle of $\Delta \theta = 0.01$ rad $= 0.6\deg$.
Fig.~\ref{fig:images} shows that the this rotation was sufficient to move off of the Bragg
peak at the center of the mirror and to the Bragg peak at the edge, indicating that there
is a 1\% gradient of the ML $D$ spacing from center to edge, consistent with the
manufacturing tolerance.

\subsection{Polarimetry Beamline Phase III}

\begin{figure}
   \centering
   \includegraphics[width=8cm]{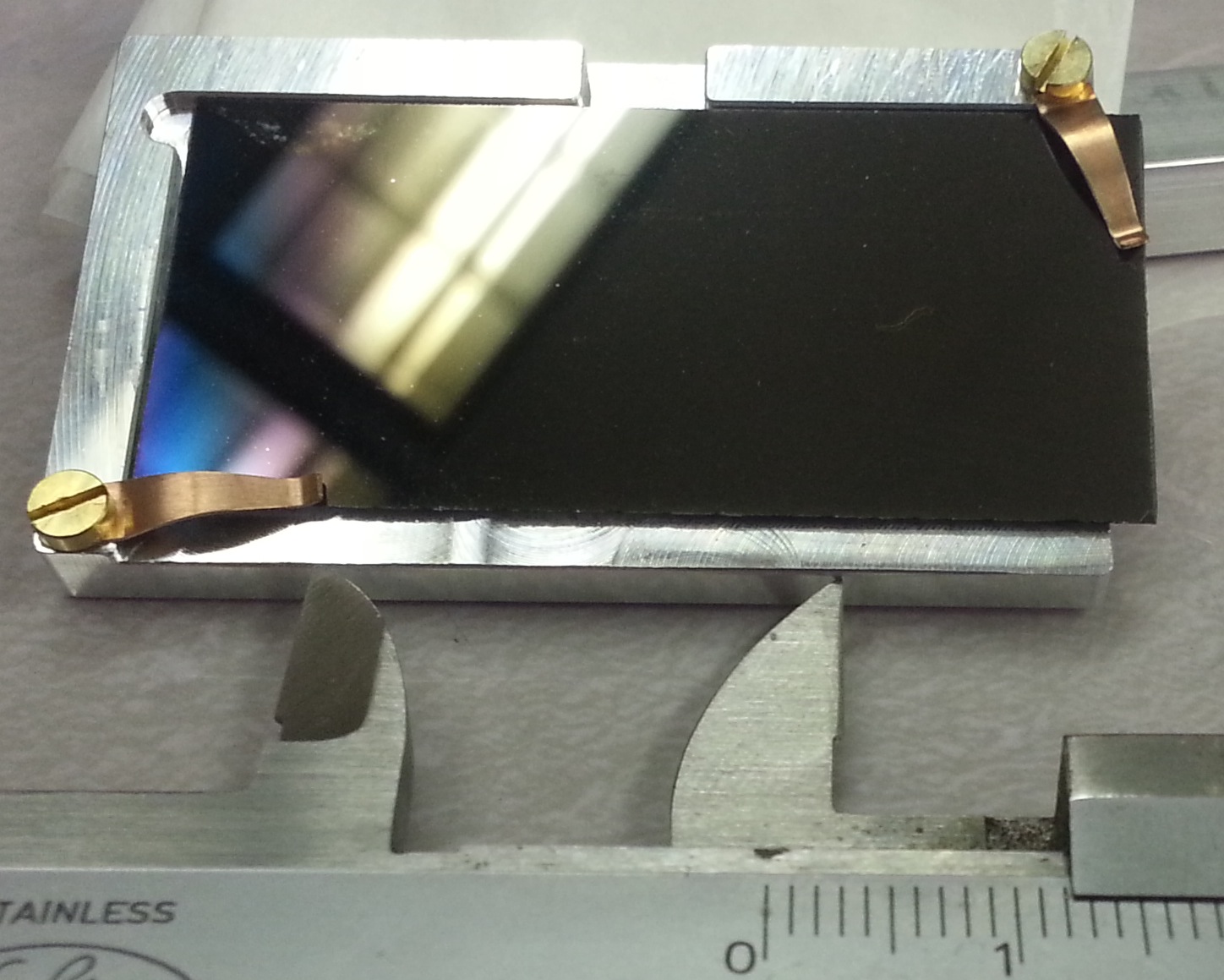}
   \includegraphics[width=8cm]{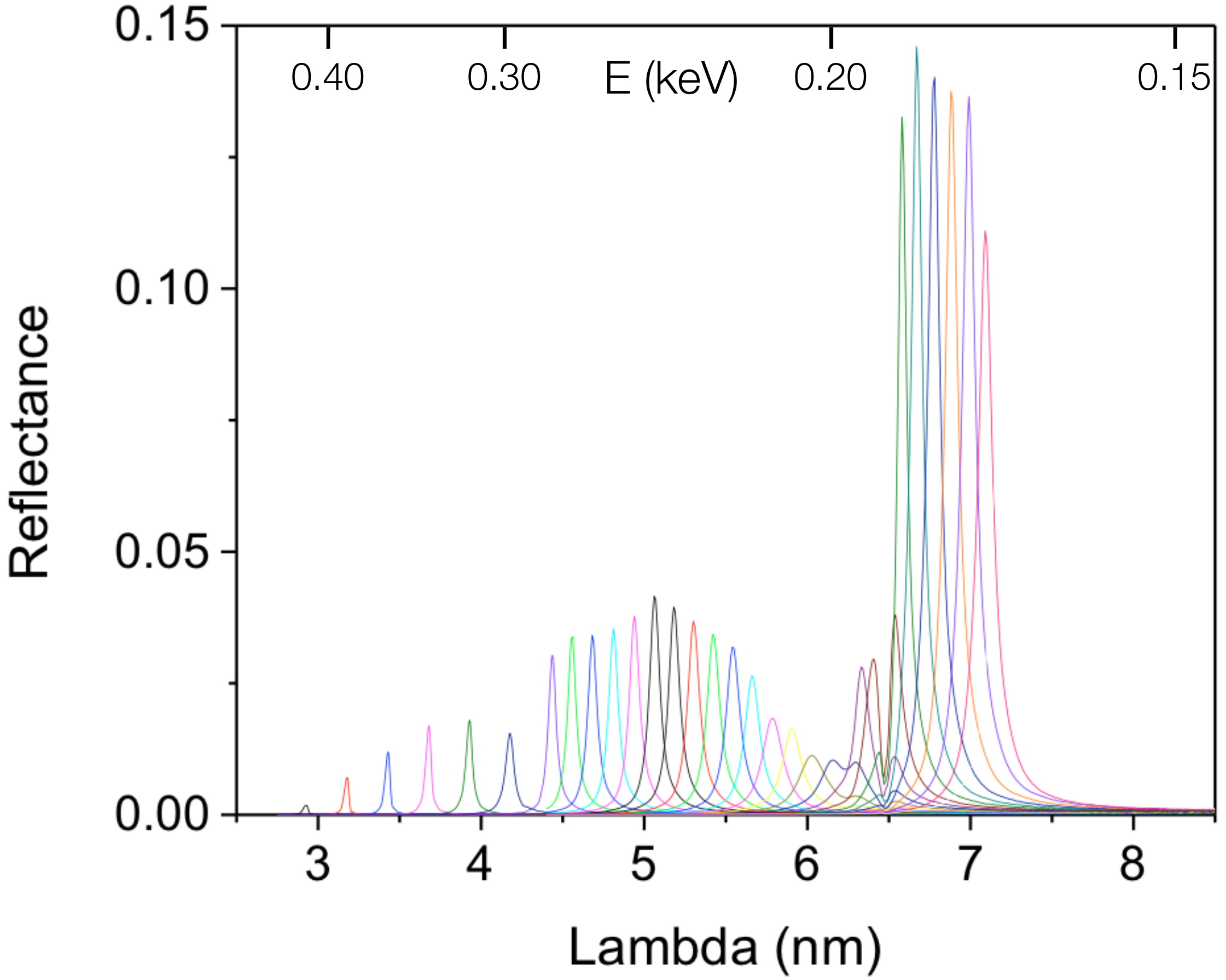}
 \caption{
 {\it Left:} A laterally graded ML coated mirror (LGML) from RXO in a holder.
  It is about 47 mm long, 23 mm wide, and 0.5 mm thick.
  The $D$ spacing increases from left to right.
  {\it Right:} Reflectivity measurements of a LGML from RXO.
  For one measurement run, the reflectivity was sampled a 2 mm spacing, starting
  below 3 nm.  In a second run, the samples were at 1 mm spacing.  The ALS
  beam consisted of about 70\% s-polarization and data were taken at a graze
  angle of 45$\deg$.
  }
\label{fig:lgml}
\end{figure}

With funding from a MIT Kavli Investment grant,
we started Phase III, where the goal is to develop, test, and install a laterally
graded ML coated mirror (LGML) in the source mirror chamber.
A pair of LGMLs were fabricated by RXO, consisting of 200 bilayers of W and
B$_4$C on highly polished Si wafers.
The $D$ spacing was varied in order
to reflect and polarize X-rays from 17\AA\ to 73\AA\ (170 to 730 eV).
Fig.~\ref{fig:lgml} shows a picture of one such LGML and the results from
a reflectivity run at the ALS.

We will soon reconfigure the source ML mirror chamber by replacing the
current single-period ML mirror with a LGML.  We have obtained a motor-controlled
rotational manipulator and will mount the LGML to its shaft so that the axis of
the shaft is centered on the long axis of the LGML's surface.  The shaft will
be controllable to 0.01$\deg$ and will be mounted on a linear bellows with
a 100 mm travel that is computer-controlled to an accuracy of 0.01 mm.
After alignment, we will calibrate the linear stage to determine the insertion
points that maximize reflectivity at a wide variety of soft X-ray emission lines
from different anodes used in the X-ray source.  For this phase, the CCD will
be returned to the position it had in Phase I.  Upon completion of Phase III,
the X-ray source will be capable of generating 100\% polarized X-rays at
a wide range of energies and rotating the polarization direction through at
least 150$\deg$.

\subsection{Polarimetry Beamline Phase IV}

\begin{figure}
  \centering
   \includegraphics[width=14cm]{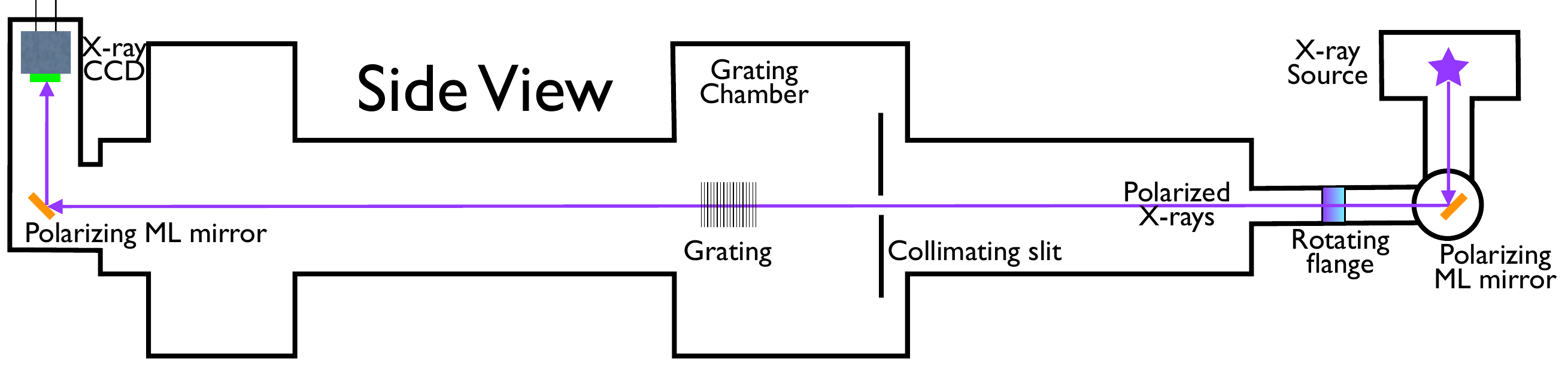}
   \includegraphics[width=2.5cm]{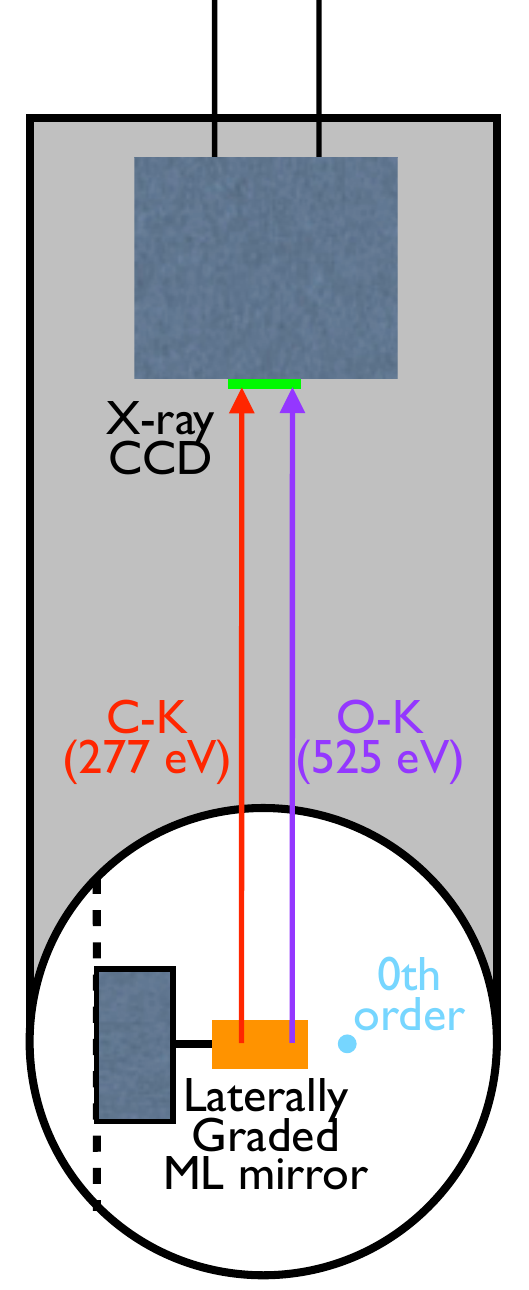}
 \caption{
 Schematic of the polarimetry beamline in Phase IV.  The configuration is the same
 as in Phase II (Fig.~\ref{fig:schematic2}) except that the polarizing ML mirrors at
 both ends of the system are replaced with LGMLs and there is a grating
 mounted in the grating chamber to disperse X-rays to specific locations on
 the detector LGML.
 {\it Left:} Side view, showing the location of the grating.
 {\it Right:} Beam view, showing that the LGML will reflect different input energies
 from different locations on its surface.
}
\label{fig:schematic4}
\end{figure}

With new funding from the NASA Astrophysics Research and Analysis (APRA) program,
we will move to Phase IV, where the goals are 1) to improve the reflectivities of LGMLs
by trying new material combinations and 2) show that a grating-LGML combination
can measure polarization over wide range of energies, thus prototyping a design
that could be used for a flight system.  See Fig.~\ref{fig:schematic4} for a
schematic of the Phase IV configuration.

\begin{figure}
   \centering
   \includegraphics[width=8cm]{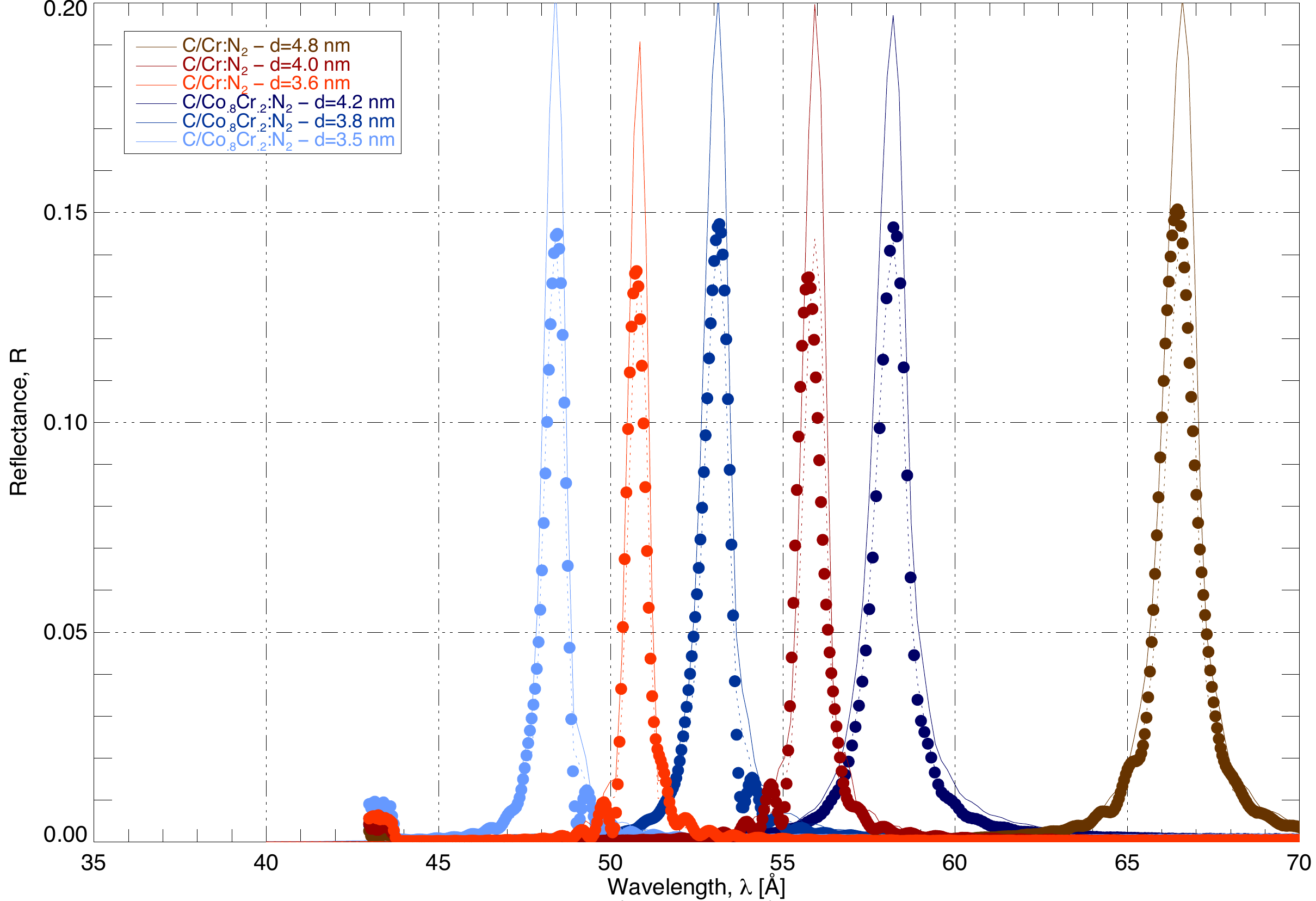}
   \includegraphics[width=8cm]{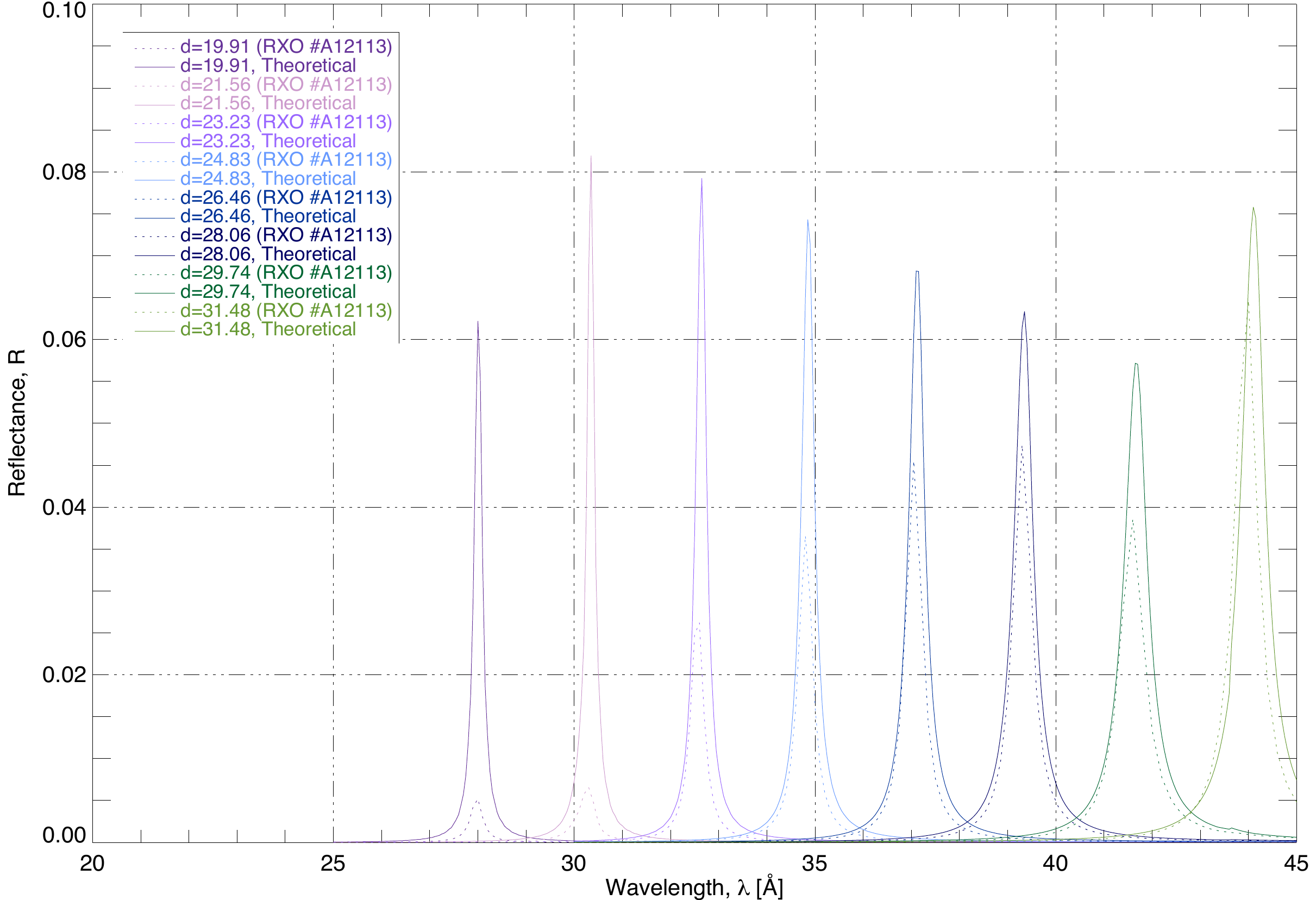}
 \caption{
 Multilayer reflectivities for single-period coatings.
  New laterally graded ML mirrors will be developed using these
  coating combinations.  {\it Left:} Several
  examples of new single-period multilayer
  coating samples that include C for use in the 44-65 \AA\ band.
  Data are shown with model fits (dotted lines) from the partially polarized ALS
  measurements.  Solid lines show reflectivity for the same models but for 100\% polarized X-rays,
  which indicate that the s-polarization reflectivities approach 20\% in this wavelength range.
  {\it Right:} Measurements from a Laterally graded ML coated sample from RXO
  are shown with dotted lines compared to theoretically achievable reflectivities.
  The coating was
  made by sputtering 200 bilayers of W and B$_4$C with N$_2$ in the chamber.  By varying
  the deposition approach, we expect to improve reflectivities to approach the theoretical
  values.
}
\label{fig:ccr}
\end{figure}

The first improvement to the LGML reflectivities will involve using different
ML compositions for specific wavelength regions.  For example, MLs with C/Cr
bilayers are known to perform better than W/B$_4$C MLs.  Figure~\ref{fig:ccr}
shows results from ALS testing of the first few single-period MLs involving
C and Cr or a CoCr alloy.  The C/CoCr MLs are slightly better and the
predicted reflectivities to s-polarization approach 20\% in the 45-65 \AA\ range.
These reflectivites are substantially better than the 2-5\% values currently
available in the W/B$_4$C LGMLs (Fig.~\ref{fig:lgml}, right).
ML coating process adjustments will be varied in order to improve the
reflectivities below 45 \AA.

The second advance in Phase IV will be to use gratings to disperse the X-rays
to the LGML in the detector chamber.  The dispersion of the grating is given by the
grating equation: $m \lambda = P sin \phi$, where $P$ is the grating period, $m$
is the grating order of interest (which we take to be $+1$) and $\phi$ is the
dispersion angle.  Defining $y$ to be the horizontal direction in Fig.~\ref{fig:schematic4} right
and defining $y = 0$ to be where the 0th order lands at the plane of the LGML,
then we match the LGML's Bragg peak to the grating dispersion by setting
$P sin \phi = P y/D_g = 2 D(y) \sin \theta = \surd 2 ~ D(y)$, where $D_g$ is
the distance from the grating to the LGML, giving $D = P y / (D_g \surd 2)$.
As long as $D(y)$ is linear, the LGML can be placed at the appropriate distance, $D_g$,
from the grating to reflect X-rays of arbitrary wavelengths, within the physical limitation
of the LGML.  The current LGMLs have $D$ gradients of 0.87 \AA/mm, which is
matchable by gratings made for the {\it Chandra} Low Energy Transmission
Grating (LETG) Spectrometer \cite{1997SPIE.3113..172P} for $D_g = 8.7$ m.  We have four LETG facets
on loan from MPE (courtesy P. Predehl) that
will be mounted in the grating chamber for this purpose.

\section{A Soft X-ray Polarizing Spectrometer}
\label{sec:instr}

\begin{figure}
    \centering
   \includegraphics[height=11cm]{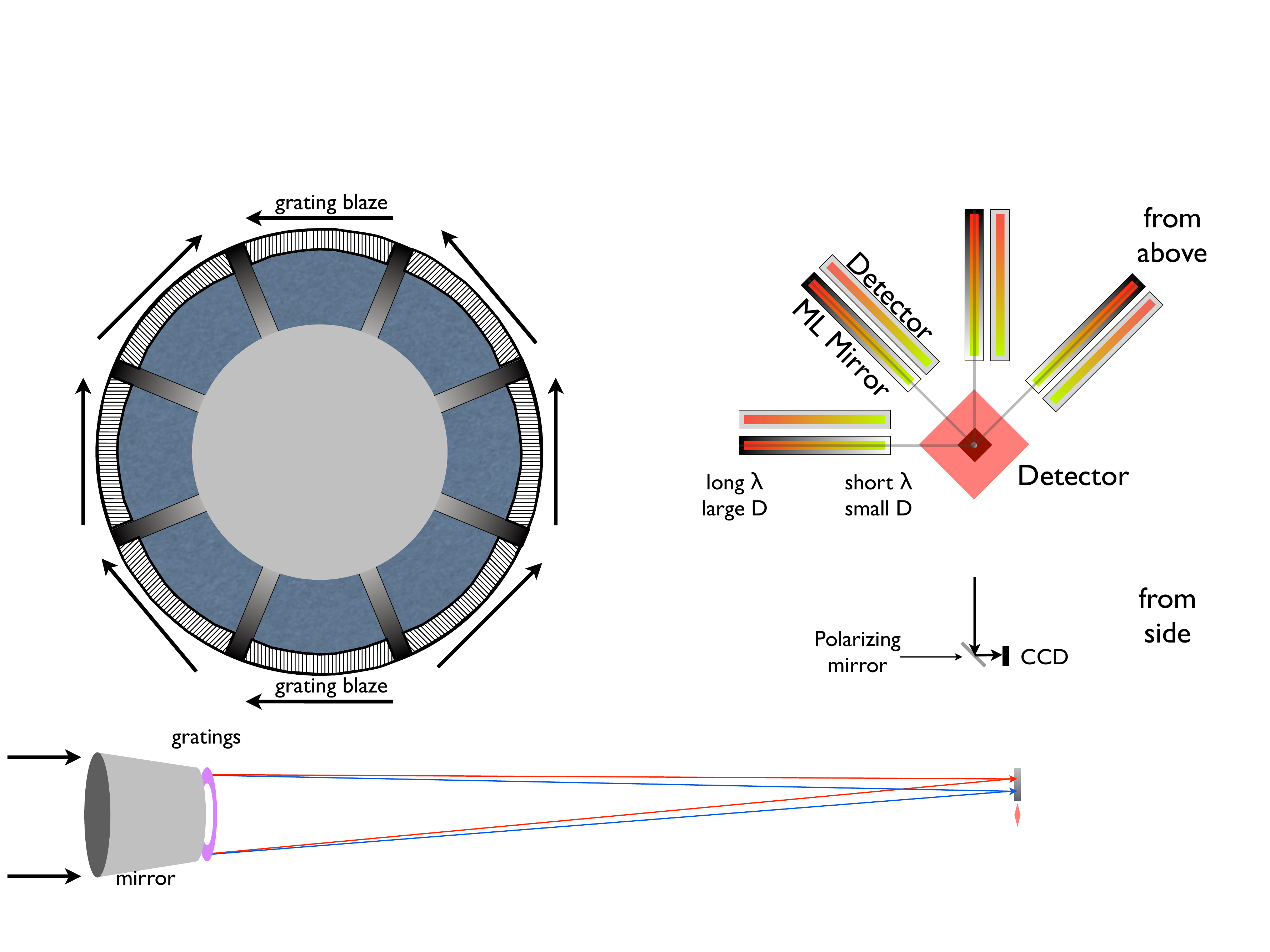}
 \caption{
 Schematic of a suborbital soft X-ray polarimeter using blazed gratings such as the
   CAT gratings under development at MIT \cite{2009SPIE.7437E..14H}.
 {\it Bottom:} The target is to the left and the dispersion by one set of gratings is shown.
 {\it Left:} View of the front aperture, where blazed gratings are oriented approximately radially
    and in sectors to improve resolution along the dispersion direction.
 {\it Right, above:} Top view of a focal plane layout that could be used for a suborbital rocket
         experiment, in the manner suggested by
	Marshall (2008 \cite{2008SPIE.7011E..63M}).
	The zeroth order is placed at the location of the gray dot so that
	the dispersed spectrum first intercepts the laterally graded multilayer mirror that is angled at
	$45\deg$ to the incoming X-rays.
 {\it Right, below:} Side view, where the dispersion is perpendicular to the plane
	of the drawing and the multilayer mirror is oriented $45\deg$ to the incoming, dispersed X-rays.
}
\label{fig:suborbital}
\end{figure}

The basic design of a polarizing spectrometer
was outlined by Marshall (2008 \cite{2008SPIE.7011E..63M}).  For this paper,
we examine the approach that could be applied to a suborbital rocket experiment.
Figure~\ref{fig:suborbital} shows a possible schematic for a suborbital mission using
blazed gratings such as the Critical Angle Transmission
(CAT) gratings under development at MIT \cite{Heilmann08,2009SPIE.7437E..14H}.
Sampling at least 3 position angles is required in order to measure
three Stokes parameters (I, Q, U) uniquely, so one would require at
least three separate detector
systems (one of which could be just for 0th order)
with accompanying multilayer-coated flats or that the rocket
rotate during the observations (which is expected anyway, to take out
systematic effects).

\begin{figure}
   \centering
   \includegraphics[width=8cm]{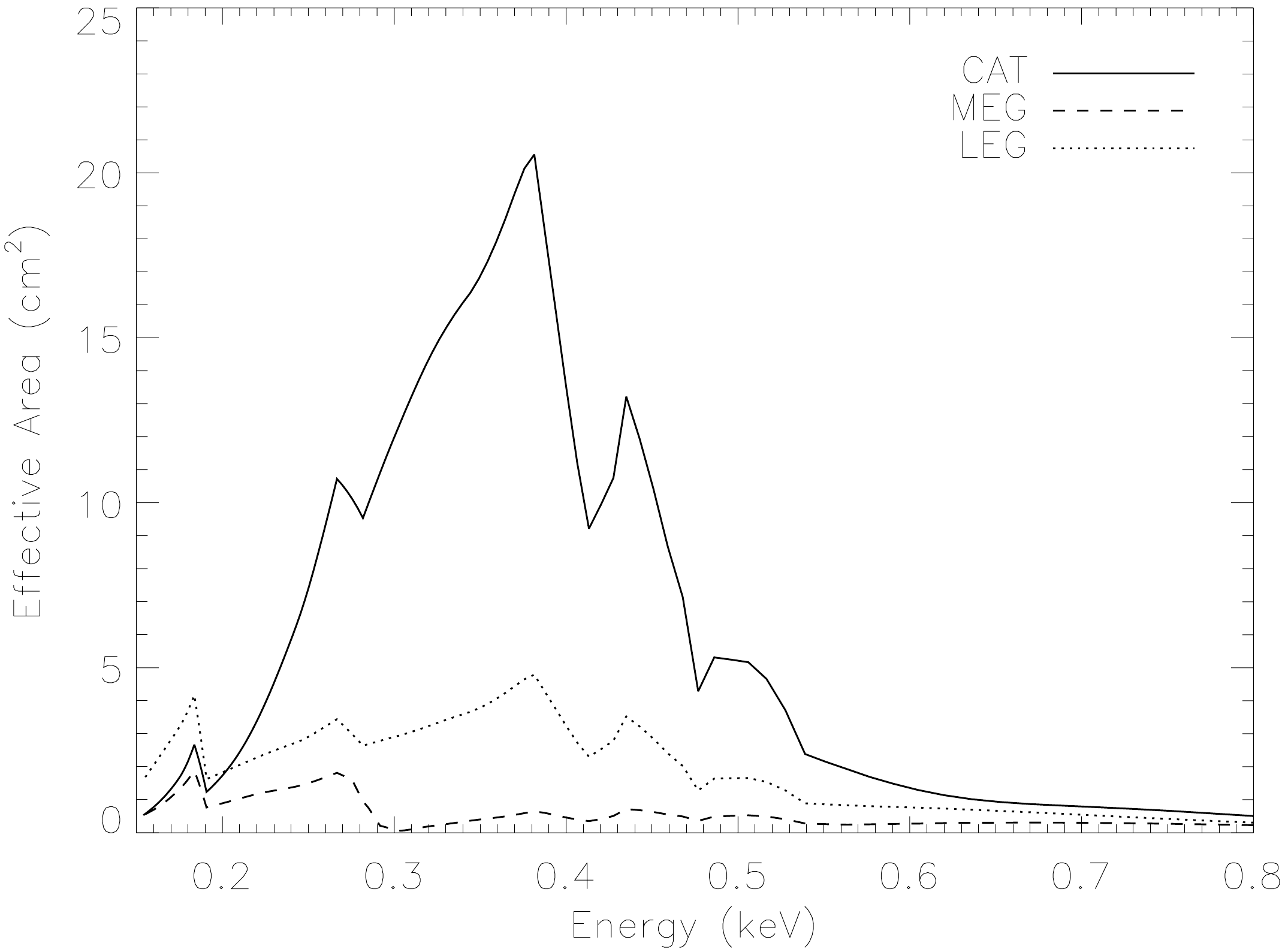}
   \includegraphics[width=8cm]{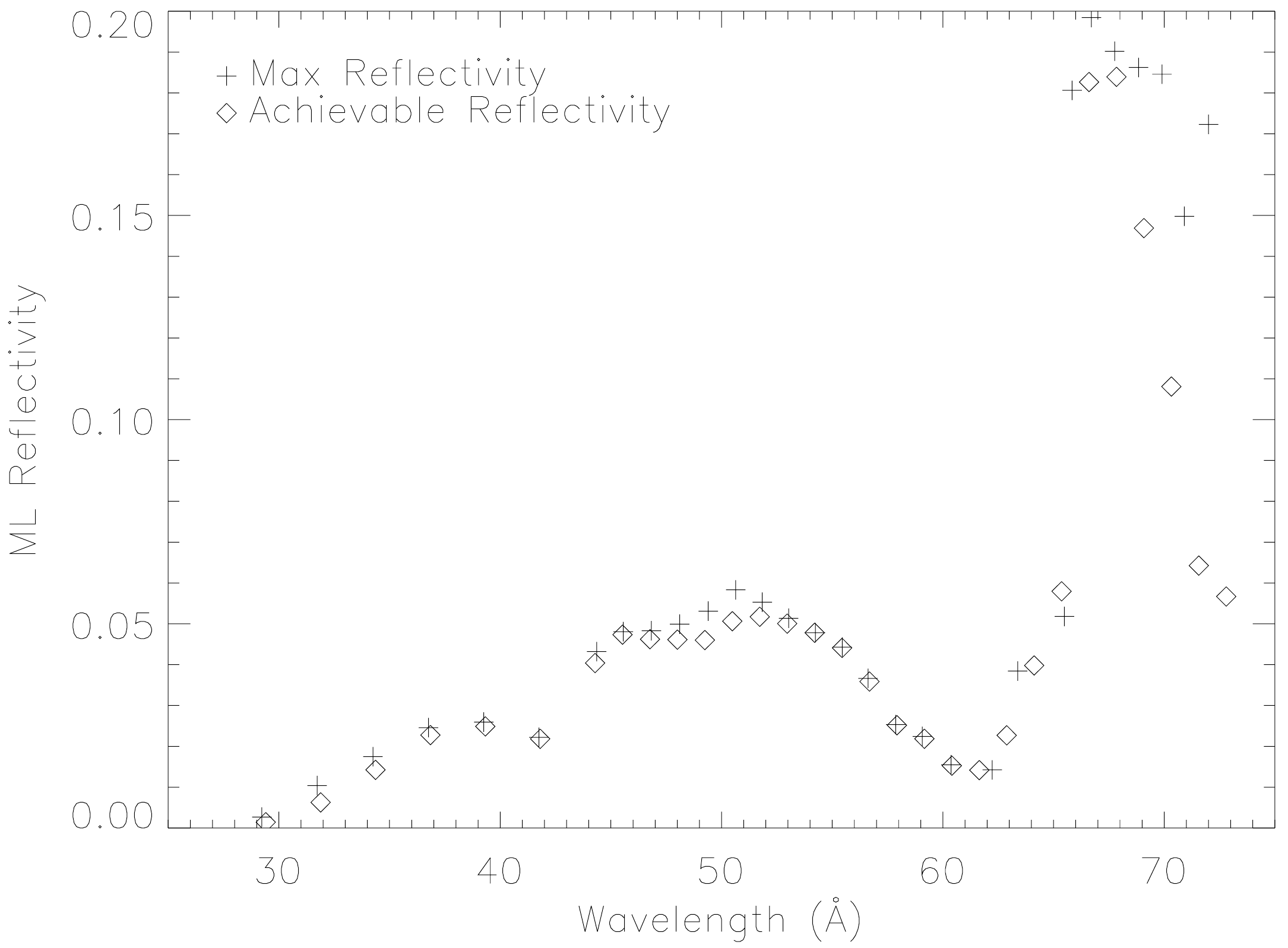}
 \caption{
 {\it Left:} Effective area of a suborbital polarimeter mission using achievable
 ML reflectivities.
  {\it Right:} Maximum reflectivities at various locations on a LGML from RXO
  (plus signs),
  compared to values obtained using a strictly linear dispersion (diamonds).
}
\label{fig:effarea}
\end{figure}

The system design consists of a mirror system with an assumed effective area of 350 cm$^2$
below 1 keV, backside-illuminated CCD detectors like those on {\em Chandra} with thin
directly deposited optical blocking filters, and CAT gratings blazed to maximize efficiency at 300 eV.
We have also computed effective areas using medium energy gratings (MEGs \cite{hetgs}) and
LETGs \cite{1997SPIE.3113..172P}.
For ML coating reflectivities, we used values that have been achieved in the lab
for single-period MLs used at 45$\deg$ and
interpolated using comparable theoretical models.
The resulting achievable effective area is shown in Fig.~\ref{fig:effarea}.

The effective area estimate can be used to predict the minimum
detectable polarization (MDP)
for a potential target.
Extragalactic sources such as the BL Lac object
Mk 421 are expected to be highly polarized in the soft X-ray band.
Mk 421 is currently the brightest BL Lac object, and a recent
{\em Chandra} LETG spectrum was readily fit by a power law
spectrum ($n_E = n_0 E^{-\Gamma} e^{-N_H \sigma[E]}$, where $\sigma[E]$ is
the energy dependent atomic cross section of the interstellar medium with
cosmic abundances and $E$ is in keV) with $n_0 = 0.2$ ph cm$^{-2}$ s$^{-1}$ keV$^{-1}$
$\Gamma = 2.7$, and an assumed cold column density of $N_H = 1.61 \times 10^{20}$ cm$^{-2}$.
In a 500 s observation of Mk 421,
this instrument could detect polarizations of 3.9\% using CAT gratings,
11.5\% using MEGs, or 6.5\% using LETGs.

LGMLs with the achievable reflectivities have not yet been
fabricated, so we recomputed the MDPs for reflectivities as measured by the ALS
for the LGML made by RXO.  Interpolating using the Bragg peak reflectivities
at the measured energies gives MDPs of 11.4\% for CAT gratings, and 16\% using LETGs.
However, we note that the Bragg peaks do not precisely correspond to the locations
of a linearly dispersed spectrum, so that the reflectivity will be somewhat smaller than
measured at some wavelengths.  See Fig.~\ref{fig:effarea} (right) for estimates of how much
the reflectivities will decrease.  We have used the observed reflectivity curves
to compute that the MDP will increase by about 10\% due to slight nonlinearity
of the actual manufactured LGML, so we expect that a suborbital flight using
the {\em existing} LGMLs would reach 18\% using existing LETGs or
improve to 12\% should CAT (or high efficiency
reflection) gratings be available.  Continued development of LGMLs are
expected to bring these MDPs down below 10\%.

\acknowledgments     

We are very grateful for the assistance and support provided by Steve Kissel
and Beverly LaMarr in providing, operating, and modifying the CCD detector system.
MIT undergraduates Keven Jenks and Kelly Kochanski helped set up and operate
the X-ray source.
We thank Regina Soufli for suggesting and facilitating the ALS measurements
of the laterally graded multilayer coatings.
Support for this work was provided by the National Aeronautics and
Space Administration through grant NNX12AH12G and by Research
Investment Grants from the MIT Kavli Institute.


\bibliography{polarimetry13}   
\bibliographystyle{spiebib}   

\end{document}